\documentstyle[12pt]{article}
\input psfig.sty
\def\nn{\noindent}

\def\Re{{\cal R \mskip-4mu \lower.1ex \hbox{\it e}\,}}
\def\Im{{\cal I \mskip-5mu \lower.1ex \hbox{\it m}\,}}
\def\ie{{\it i.e.}}

\def\etal{{\it et al.}}
\def\ibid{{\it ibid}.}
\def\sub#1{_{\lower.25ex\hbox{$\scriptstyle#1$}}}

\def\to{\rightarrow}

\def\subw{_{\rm w}}
\def\mh{\ifmmode m\sbl H \else $m\sbl H$\fi}
\def\mch{\ifmmode m_{H^\pm} \else $m_{H^\pm}$\fi}
\def\mt{\ifmmode m_t\else $m_t$\fi}
\def\mc{\ifmmode m_c\else $m_c$\fi}
\def\mz{\ifmmode M_Z\else $M_Z$\fi}
\def\mw{\ifmmode M_W\else $M_W$\fi}
\def\mws{\ifmmode M_W^2 \else $M_W^2$\fi}
\def\mhs{\ifmmode m_H^2 \else $m_H^2$\fi}   
\def\mzs{\ifmmode M_Z^2 \else $M_Z^2$\fi}
\def\mts{\ifmmode m_t^2 \else $m_t^2$\fi}
\def\mcs{\ifmmode m_c^2 \else $m_c^2$\fi}
\def\mchs{\ifmmode m_{H^\pm}^2 \else $m_{H^\pm}^2$\fi}
\def\ztwo{\ifmmode Z_2\else $Z_2$\fi}
\def\zone{\ifmmode Z_1\else $Z_1$\fi}
\def\mtwo{\ifmmode M_2\else $M_2$\fi}
\def\mone{\ifmmode M_1\else $M_1$\fi}
\def\tb{\ifmmode \tan\beta \else $\tan\beta$\fi}
\def\xw{\ifmmode x\subw\else $x\subw$\fi}
\def\ch{\ifmmode H^\pm \else $H^\pm$\fi}
\def\lum{\ifmmode {\cal L}\else ${\cal L}$\fi}
\def\inpb{\ifmmode {\rm pb}^{-1}\else ${\rm pb}^{-1}$\fi}
\def\infb{\ifmmode {\rm fb}^{-1}\else ${\rm fb}^{-1}$\fi}
\def\epem{\ifmmode e^+e^-\else $e^+e^-$\fi}
\def\ppb{\ifmmode \bar pp\else $\bar pp$\fi}
\def\mpl{\ifmmode \overline M_{Pl}\else $\bar M_{Pl}$\fi}

\newskip\zatskip \zatskip=0pt plus0pt minus0pt
\def\matth{\mathsurround=0pt}
\def\lsim{\mathrel{\mathpalette\atversim<}}
\def\gsim{\mathrel{\mathpalette\atversim>}}
\def\atversim#1#2{\lower0.7ex\vbox{\baselineskip\zatskip\lineskip\zatskip
  \lineskiplimit 0pt\ialign{$\matth#1\hfil##\hfil$\crcr#2\crcr\sim\crcr}}}

\def\be{\begin{equation}}
\def\ee{\end{equation}}
\def\bea{\begin{eqnarray}}
\def\eea{\end{eqnarray}}
\renewcommand{\thefootnote}{\fnsymbol{footnote}}

\begin{document} \begin{titlepage} 
\rightline{\vbox{\halign{&#\hfil\cr
&SLAC-PUB-8241\cr
&September 1999\cr}}}
\vspace{1in} 
\begin{center}

{\Large\bf
Warped Phenomenology}
\footnote{Work supported by the Department of 
Energy, Contract DE-AC03-76SF00515}
\medskip

\normalsize 
{\large H. Davoudiasl, J.L. Hewett and T.G. Rizzo \\}
\vskip .3cm
Stanford Linear Accelerator Center \\
Stanford CA 94309, USA\\
\vskip .3cm

\end{center}

\begin{abstract} 

We explore the phenomenology associated with the recently proposed localized
gravity model of 
Randall and Sundrum where gravity propagates in a 5-dimensional non-factorizable
geometry and generates the 4-dimensional weak-Planck scale hierarchy by an
exponential function of the compactification radius, called a warp factor.
The Kaluza-Klein tower of gravitons which emerge in this scenario have 
strikingly different properties than in the factorizable case with large 
extra dimensions.  We derive the form of the graviton tower interactions 
with the Standard Model fields and examine their direct production in 
Drell-Yan and dijet events at the Tevatron and LHC as well as the KK spectrum 
line-shape at high-energy linear \epem\ colliders.   In the case where the
first KK excitation is observed, we outline the procedure to uniquely determine 
the parameters of this scenario.  We also investigate the effect of KK tower 
exchanges in contact interaction searches.  We find that present experiments 
can place meaningful constraints on the parameters of this model.

\end{abstract}

\renewcommand{\thefootnote}{\arabic{footnote}} \end{titlepage} 


The large disparity between the electroweak and apparent fundamental scale 
of gravity, known as the hierarchy problem,  is a primary mystery 
of particle physics.  Traditionally, new symmetries, particles, or interactions
have been introduced at the electroweak scale to stabilize this hierarchy.
However, it is possible that our 4-dimensional vision of gravity does not
represent the full theory and that the observed value of the Planck scale,
$M_{Pl}$ is not truly fundamental.  A scenario of this type due to Arkani-Hamed,
Dimopoulos, and Dvali\cite{nima} (ADD) proposes
the existence of $n$ additional compact dimensions and relates the
fundamental $4+n$ dimensional Planck scale, $M$, to our effective 4-dimensional
value through the volume of the compactified dimensions, $M^2_{Pl}=V_nM^{2+n}$.
Setting $M\sim$ TeV to remove the above hierarchy necessitates a large
size for the extra dimensions with a compactification scale of $\mu_c=1/r_c
\sim$ eV$-$MeV for $n=2-7$.  This, unfortunately, introduces another hierarchy
between $\mu_c$ and $M$, which must somehow be stabilized.  Nonetheless,
this scenario has received much attention as it affords concrete 
phenomenological tests.
Since it is experimentally determined that the
Standard Model (SM) fields do not feel the effects of additional dimensions of 
this size, they are confined to a wall, or 3-brane, while gravity is 
allowed to propagate freely in the full higher-dimensional space, or bulk.  
Kaluza-Klein (KK) towers of gravitons, which can interact with the wall fields, 
result from compactification of the bulk.  The coupling of each KK excitation
is $M_{Pl}$ suppressed, however the mode spacing is determined by $\mu_c$ and 
is thus very small compared to typical collider energies.  This allows the 
summation over an enormous number of KK states which can be exchanged or 
emitted in a physical process, thereby reducing the summed suppression from 
$1/M_{Pl}$ to $1/M$, or $\sim$TeV$^{-1}$. This has resulted in
a vast array of phenomenological\cite{pheno} and astrophysical\cite{astro} 
studies with present collider data bounding $M\gsim 1$ TeV for all $n$ and 
Supernova 1987A cooling and $\gamma$ ray flux constraints setting 
$M\gsim 50-110$ TeV for $n=2$ only.

An alternative higher dimensional scenario has recently been proposed by
Randall and Sundrum\cite{rs1} (RS), where the hierarchy is generated by an
exponential function of the compactification radius, called a warp factor.
They assume a 5-dimensional non-factorizable geometry, based on a slice
of $AdS_5$ spacetime.  Two 3-branes, one being visible with the other being
hidden, with opposite tensions reside at
$S_1/Z_2$ orbifold fixed points, taken to be $\phi=0,\pi$, where $\phi$ is
the angular coordinate parameterizing the extra dimension.  The 
solution to Einstein's equations for this configuration, maintaining
4-dimensional Poincare invariance, is given by the 5-dimensional metric
\be
ds^2=e^{-2\sigma(\phi)}\eta_{\mu\nu}dx^\mu dx^\nu+r_c^2d\phi^2 \,,
\label{5metric}
\ee
where the Greek indices run over ordinary 4-dimensional spacetime, 
$\sigma(\phi)=kr_c|\phi|$ with $r_c$ being the compactification radius of the
extra dimension, and $0\leq |\phi|\leq\pi$.  Here $k$ is a scale of
order the Planck scale and relates the 5-dimensional Planck scale $M$ to the 
cosmological constant.  Similar configurations have also been found to arise in 
M/string-theory\cite{strings}.  An extension of this scenario where the
higher dimensional space is non-compact, \ie, $r_c\to \infty$, is discussed
in Ref. \cite{rs2} and several aspects of this and related ideas have been
investigated in Ref. \cite{morestuff}.  Examination of the action in the 
4-dimensional effective theory in the RS scenario yields\cite{rs1}
\be
\mpl^2={M^3\over k}(1-e^{-2kr_c\pi}) 
\label{mpl}
\ee
for the reduced effective 4-D Planck scale.  
Assuming that we live on the 3-brane located at $|\phi|=\pi$, it is found
that a field on this brane with the fundamental mass
parameter $m_0$ will appear to have the physical mass $m=e^{-kr_c\pi}m_0$.
TeV scales are thus generated from fundamental scales of order $M_{Pl}$
via a geometrical exponential factor and the observed scale hierarchy is
reproduced if $kr_c\simeq 12$.  Hence, due to the exponential nature of the
warp factor, no additional large hierarchies are generated.  In fact, it has 
been demonstrated\cite{gw2} that the 
size of $\mu_c$ in this scenario can be stabilized without fine tuning of 
parameters, making this theory very attractive.

The graviton KK spectrum is quite different in this scenario than in the 
case with factorizable geometry, resulting in a distinctive phenomenology.
As we will see below, the masses and couplings of each individual KK excitation
are determined by the scale $\Lambda_\pi=\mpl e^{-kr_c\pi}\sim$ TeV.
This implies that these KK states can be separately produced on resonance with
observable rates at colliders up to the kinematic limit.   We will examine 
the cases of KK graviton production in Drell-Yan and dijet events at hadron
colliders as well as the KK spectrum line-shape at high-energy linear \epem\
colliders.  In the circumstance where a resonance is observed, we outline the 
procedure to be employed in order to uniquely determine the parameters of this
model.  In the case where no direct production is observed, we compute
the bounds on the parameter space in the contact interaction limit.  We find
that data from present accelerators already place meaningful constraints on
the parameter space of this scenario.
The phenomenology of these KK gravitons  is similar in spirit to the 
production of traditional KK excitations of the SM gauge fields\cite{ant}, 
but differs in detail because of the form of the KK wavefunction due to the
non-factorizable metric and their spin.

We now calculate the mass spectrum and couplings of the graviton 
KK modes in the effective 4-dimensional theory on the 3-brane at $\phi = \pi$.  
The starting point is the 5-dimensional Einstein's equation for the RS
configuration, which is given in Ref. \cite{rs1}.  We parameterize the tensor 
fluctuations $h_{\alpha \beta}$ by taking a linear expansion of the flat 
metric about its Minkowski value, $\hat{G}_{\alpha \beta} = e^{-2 \sigma} 
\left(\eta_{\alpha \beta} + \kappa^* \,  h_{\alpha \beta}\right)$,
where $\kappa^*$ is an expansion parameter.  In order to obtain the mass 
spectrum of the tensor fluctuations, we consider the 4-dimensional
$\alpha \beta$ components of Einstein's equation with the replacement 
$G_{\alpha \beta} \to \hat{G}_{\alpha \beta}$,
keeping terms up to $\cal{O}(\kappa^*)$.  We work in the gauge with 
$\partial^\alpha h_{\alpha\beta} = h^\alpha_\alpha = 0$.  Upon compactification
the graviton field $h_{\alpha \beta}$ is expanded into a KK tower
\begin{equation}
h_{\alpha \beta} (x, \phi) = \sum_{n = 0}^\infty h_{\alpha \beta}^{(n)} (x) \, 
\frac{\chi^{(n)}(\phi)}{\sqrt{r_c}}\,,
\label{hKK}
\end{equation} 
where the $h_{\alpha \beta}^{(n)} (x)$ correspond to the KK modes of the 
graviton on the background of Minkowski space on the 3-brane.  In a gauge where 
$\eta^{\alpha \beta} \partial_\alpha h^{(n)}_{\beta \gamma} = 
\eta^{\alpha \beta} h_{\alpha \beta}^{(n)} = 0$, the equation of motion of 
$h_{\alpha \beta}^{(n)}$ is given by
\begin{equation}
\left(\eta^{\alpha \beta} \partial_\alpha \partial_\beta - m_n^2\right) 
h_{\mu\nu}^{(n)} (x) = 0\,,
\label{eom}
\end{equation}
corresponding to the states with masses $m_n \geq 0$.  
Using the KK expansion (\ref{hKK}) for $h_{\alpha \beta}$ in 
$\hat{G}_{\alpha \beta}$, Einstein's equation  in conjunction with the
above equation of motion yields the following differential equation for 
$\chi^{(n)}(\phi)$
\begin{equation}
\frac{- 1}{r_c^2} \frac{d}{d \phi} \left(e^{- 4 \sigma} \frac{d \chi^{(n)}}{d 
\phi}\right) = m_n^2 \, e^{- 2 \sigma} \chi^{(n)}\,.        
\label{diffeq}
\end{equation}
The orthonormality condition for $\chi^{(n)}$ is found to be
$\int_{- \pi}^{\pi}d\phi\, \, e^{-2\sigma}\chi^{(m)}\chi^{(n)}=\delta_{m n}$.
In deriving Eq. (\ref{diffeq}), we have used $\left(d\sigma/d\phi\right)^2= 
(kr_c)^2$ and $d^2\sigma/d\phi^2=2kr_c\left[\delta(\phi)-\delta 
(\phi-\pi)\right]$, as required by the orbifold symmetry for 
$\phi\in [- \pi,\pi]$ \cite{rs1}.  The solutions for $\chi^{(n)}$ are then
given by\cite{gw1}
\begin{equation}
\chi^{(n)}(\phi) = \frac{e^{2 \sigma}}{N_n} \left[J_2 (z_n) + \alpha_n \, Y_2 
(z_n)\right], 
\label{chi}
\end{equation}
where $J_2$ and $Y_2$ are Bessel functions of order 2, $z_n (\phi) = 
m_n e^{\sigma (\phi)}/k$, $N_n$ represents the wavefunction normalization,
and $\alpha_n$ are constant coefficients. 

Defining $x_n \equiv z_n(\pi)$, and working in the limit that $m_n/k\ll 1$ 
and $e^{kr_c\pi}\gg 1$, the requirement that
the first derivative of $\chi^{(n)}$ be continuous at the orbifold fixed 
points yields 
\be
\alpha_n\sim x_n^2e^{-2kr_c\pi}\,,\quad\quad\quad \mbox{and}\quad\quad\quad
J_1(x_n)=0\,, 
\ee
so that the $x_n$ are simply roots of the Bessel function of order 1.  
Note that the masses of the graviton KK excitations, given by 
$m_n=kx_ne^{-kr_c\pi}$, are dependent on the roots of $J_1$ and are not
equally spaced, contrasted to most KK models with one extra dimension.
For $x_n\ll e^{kr_c\pi}$, 
we see that $\alpha_n\ll 1$, and hence $Y_2(z_n)$ can be neglected compared to 
$J_2(z_n)$ in Eq. (\ref{chi}).  We thus obtain for the normalization
\begin{equation}
N_n\simeq\frac{e^{kr_c\pi}}{\sqrt{kr_c}}\, J_2(x_n) \, \, ; \, \, n > 0\,,
\label{Nn}
\end{equation}
and the normalization of the zero mode is simply $N_0=1/\sqrt{kr_c}$.

Having found the solutions for $\chi^{(n)}$, we can now derive the interactions
of $h_{\alpha \beta}^{(n)}$ with the matter fields on the 3-brane.  
Starting with the 5-dimensional action and imposing 
the constraint that we live on the brane at $\phi=\pi$, we find the usual 
form of the interaction Lagrangian in the 4-dimensional effective theory,
\be
{\cal L}=-\, {1\over M^{3/2}}T^{\alpha\beta}(x)h_{\alpha\beta}(x,\phi=\pi)\,,
\ee
where $T_{\alpha\beta}(x)$ is the symmetric conserved Minkowski space 
energy-momentum tensor of the matter fields and we have used the definition
$\kappa^* = 2/M^{3/2}$.  Expanding the graviton field into the KK states 
of Eq. (\ref{hKK})  and using the above normalization in Eq. (\ref{Nn}) 
for $\chi^{(n)}(\phi)$ we find via Eq. (2)
\be
{\cal L} = - {1\over\mpl}T^{\alpha\beta}(x)h^{(0)}_{\alpha\beta}(x)-
{1\over\Lambda_\pi}T^{\alpha\beta}(x)\sum_{n=1}^\infty 
h^{(n)}_{\alpha\beta}(x)\,.
\label{effL}
\ee
Here we see that the zero mode separates from the sum and couples with the usual
4-dimensional strength, $\mpl^{- 1}$, however, all the 
massive KK states are only suppressed by $\Lambda_\pi^{- 1}$, where we find
that $\Lambda_\pi = e^{- kr_c\pi} \mpl$, which is of order  the weak 
scale.

Our calculations have been performed with the assumption $k < M$ with
$M\sim\mpl$, so that the 5-dimensional curvature is small compared to $M$ 
and the solution for the bulk metric can be trusted\cite{rs1}.  This implies
that the ratio $k/\mpl$ cannot be too large and we take $k/\mpl\leq 1$ in our 
analysis below.  As we will see, the value of this ratio is central to the
phenomenological investigation of this model.  In order to get a feel for the
natural size of this parameter, we perform a simple estimate using string 
theoretic arguments.  The string scale $M_s$ can be related\cite{pol2} to 
$\mpl$ in 4-dimensional heterotic string theories by $M_s\sim g_{_{YM}}\mpl$,
where $g_{_{YM}}$ is the 4-dimensional Yang-Mills gauge coupling constant, 
and the tension $\tau_3$ of a $D$ 3-brane is given by 
\begin{equation}
\tau_3 = \frac{M_s^4}{g\, (2\pi)^3}\,, 
\label{tau3}
\end{equation}
where $g$ is the string coupling constant.  For $g_{_{YM}}\sim 0.7$ and 
$g\sim 1$, we find $\tau_3\sim 10^{- 3}\, \mpl^4$.  
In the RS scenario, the magnitude of the 3-brane tension is given by
$V = 24\, \mpl^2 k^2$.  Requiring that $V = \tau_3$, suggests
\begin{equation}
\frac{k}{\mpl} \sim 10^{- 2}.
\label{kmag}
\end{equation}
We take the range $0.01 \leq k/\mpl \leq 1$ in our phenomenological
analysis, however, the above discussion suggests that string theoretic 
and curvature
considerations favor the lower end of this range.  We note that recent
work\cite{ddg} on gauge unification in a modified RS scenario also favors
smaller values for this ratio.

\begin{figure}[htbp]
\centerline{
\psfig{figure=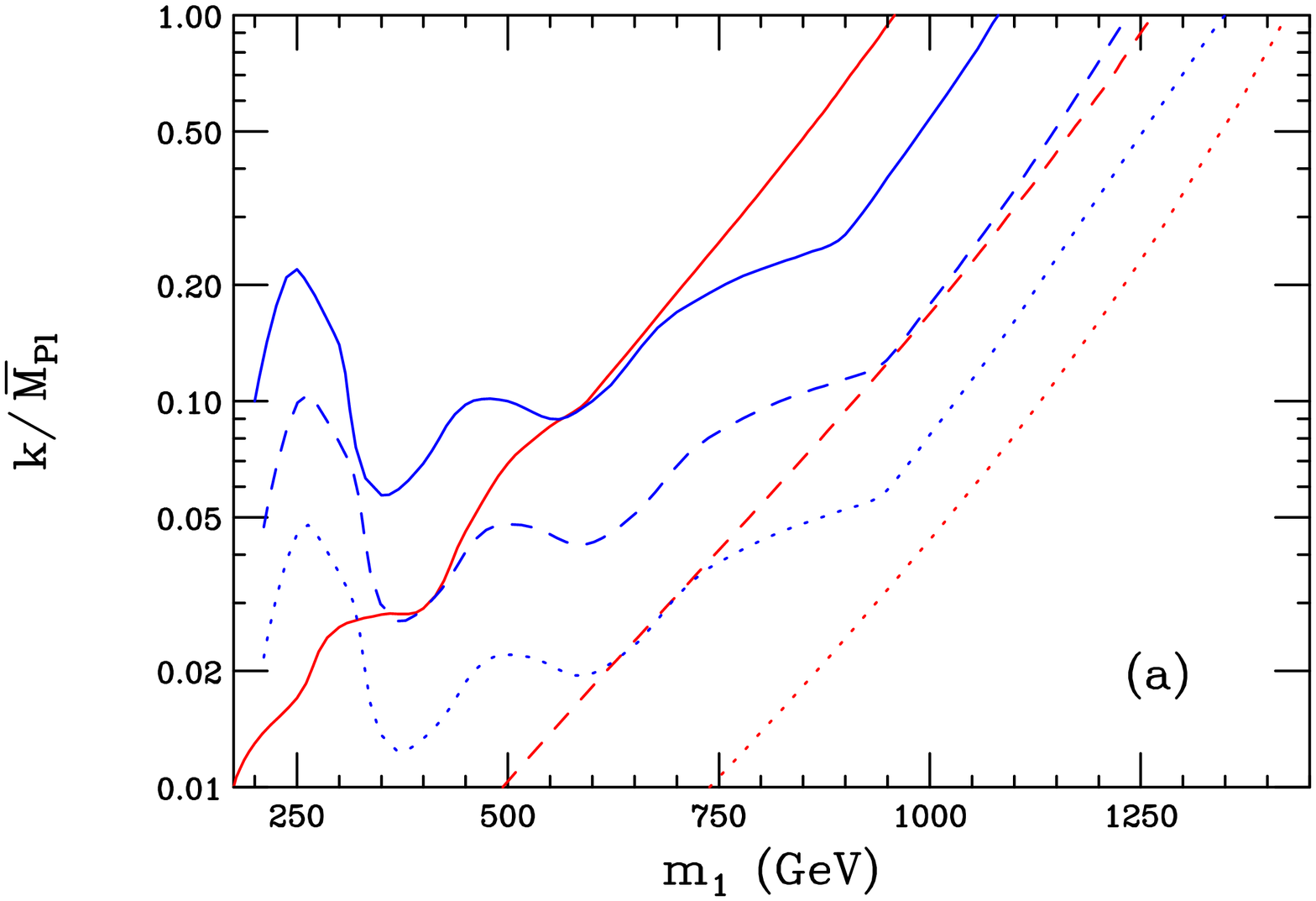,height=8.cm,width=12cm,angle=90}}
\vspace*{0.25cm}
\centerline{
\psfig{figure=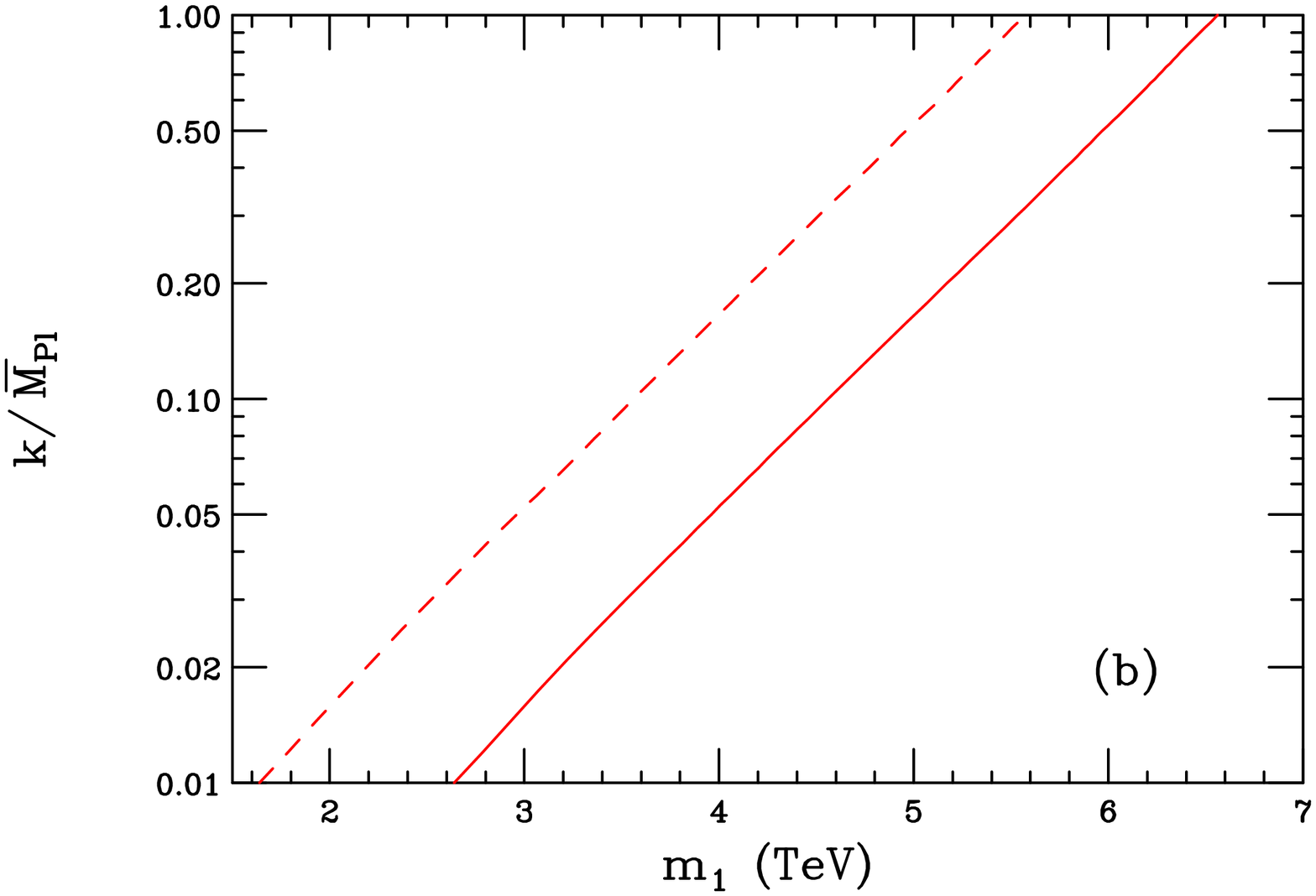,height=8.cm,width=12cm,angle=90}}
\vspace*{0.25cm}
\caption{Exclusion regions for resonance production of the first KK graviton
excitation in (a) the Drell-Yan (corresponding to the diagonal lines)
and dijet (represented by the bumpy curves) channels at the Tevatron and (b)
Drell-Yan production at the LHC.  (a) The solid curves represent the results
for Run I, while the dashed, dotted curves correspond to Run II with 2, 30
\infb\ of integrated luminosity, respectively.  
(b) The dashed, solid curves correspond to
10, 100 \infb\ . The excluded region lies above and to the left of the curves.}
\label{bumps}
\end{figure}

Constraints on the parameters of this model can be obtained by direct 
collider searches for the first graviton excitation at the Tevatron or LHC. 
The cleanest signal for graviton resonance 
production will be either an excess in 
Drell-Yan events, $q\bar q,gg \to G^{(1)} \to \ell^+\ell^-$ (in analogy to
searches for extra neutral gauge bosons), or in the dijet channel, 
$q\bar q,gg \to G^{(1)} \to q\bar q,gg$.  
Note that gluon-gluon initiated processes 
now contribute to Drell-Yan production.  This differs from the ADD scheme
where individual resonances associated with graviton exchange are not
observable due to the tiny mode spacing.
Using the above Lagrangian 
(\ref{effL}), the production cross section, decay widths, and branching 
fractions relevant for graviton production can be obtained in a 
straightforward manner.  We assume that the first excitation only decays
into SM states, so that for a fixed value of the first graviton excitation 
mass, $m_1$, the value of $k/\mpl$ completely determines all of the above 
quantities.  In fact, the total width is found to be proportional to
$(k/\mpl)^2$.  Keeping in mind that theoretic
arguments favor a smaller value for this parameter, and to get a handle on the 
possible constraints that arise from these channels, we employ
the narrow width approximation.  This is strictly valid only for 
values of $k/\mpl\lsim 0.3$ but well approximates the true search reach 
obtained via a more complete analysis\cite{dhr}.  We then compare our results 
with the existing Tevatron bounds{\cite{tev}}.  The lack 
of any signal for a new resonance 
in either the Drell-Yan or dijet channel in the data 
then provides a constraint on $k/\mpl$ for any given value of $m_1$ as shown in 
Fig. \ref{bumps}(a).  We also perform a
similar analysis to estimate the future $95\%$ C.L. parameter
exclusion regions at both Run II at the Tevatron and at the LHC under the 
assumption that no signal is found; these results are displayed in 
Figs. \ref{bumps}(a) and (b).  The dijet constraints for Run II were estimated 
by a simple luminosity (and $\sqrt s$) rescaling of the published Run I results.
Note that the Drell-Yan and dijet channels play complementary roles at the
Tevatron in obtaining these limits.  We expect a dijet search at the LHC to
yield poor results due to the large QCD background at this higher
center-of-mass energy.

\nn

The discovery of the first graviton excitation as a resonance at a collider 
will immediately allow the determination of all of the fundamental model 
parameters through measurements of its mass and width, $m_1$ and $\Gamma_1$, 
respectively. To demonstrate this, we make use of the two relations 
$\Lambda_\pi=m_1\mpl/kx_1$ and $\Gamma_1=\rho m_1x_1^2(k/\mpl)^2$, where 
$x_1$ is the first non-zero root of the $J_1$ Bessel function and $\rho$ is 
a constant which depends on the number of open decay channels; it is fixed 
provided we assume that the graviton decays only to SM fields. Using these 
relations we immediately find that $r_c=-\log[m_1/kx_1]/k\pi$ with
$k=\mpl[\Gamma_1/m_1\rho x_1^2]^{1/2}$. 
In addition, the spin-2 nature of the graviton can be determined via
angular distributions of its decay products.

\begin{figure}[htbp]
\centerline{
\psfig{figure=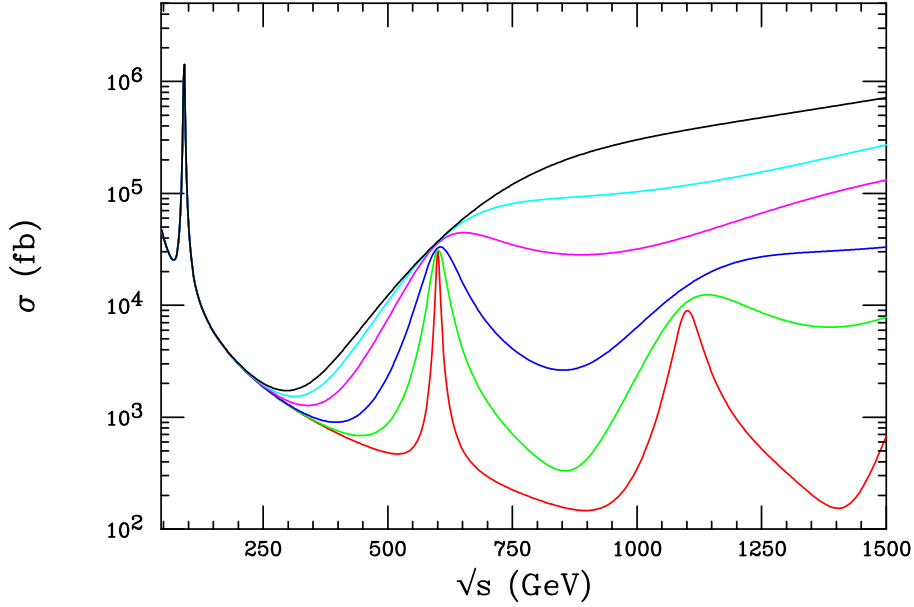,height=8.cm,width=12cm,angle=90}}
\vspace*{0.25cm}
\caption{The cross section for $\epem\to\mu^+\mu^-$ including the exchange of
a tower of KK gravitons, taking the mass of the first mode to be 600 GeV,
as a function of $\sqrt s$.  From top to bottom the curves correspond to
$k/\mpl=1.0,\, 0.7,\, 0.5,\, 0.3,\, 0.2,\, 0.1$.}
\label{lineshape}
\end{figure}

To exhibit how the tower of graviton excitations may appear at a collider, 
Fig. \ref{lineshape} displays the cross section for $e^+e^- \to \mu^+\mu^-$
as a function of $\sqrt s$, 
assuming $m_1=600$ GeV and taking various values of $k/\mpl$ for purposes of 
demonstration. We see that for small values of $k/\mpl$ the gravitons appear 
as ever widening peaks and are almost regularly spaced, with the widths and 
the spacing both being dependent on successive roots of $J_1$.
However, as $k/\mpl$ grows, the peaks become too wide to be  
identified as true resonances and the classic KK signature of successive 
peaks becomes lost. Instead, it would appear experimentally that there is an 
overall large enhancement of the 
cross section, similar to what might be expected from a contact interaction. One 
may worry that at some point the cross section may grow so large as to 
violate the partial wave unitarity bound of $\sigma_U=20\pi/s$, which is 
appropriate\cite{jimbo} to the case of initial and final fermion
states with helicity 
of 1.  However, even for values of $k/\mpl$ as large as unity we find that
unitarity will not be violated until $\sqrt s$ is at least several TeV.

\nn

\begin{figure}[htbp]
\centerline{
\psfig{figure=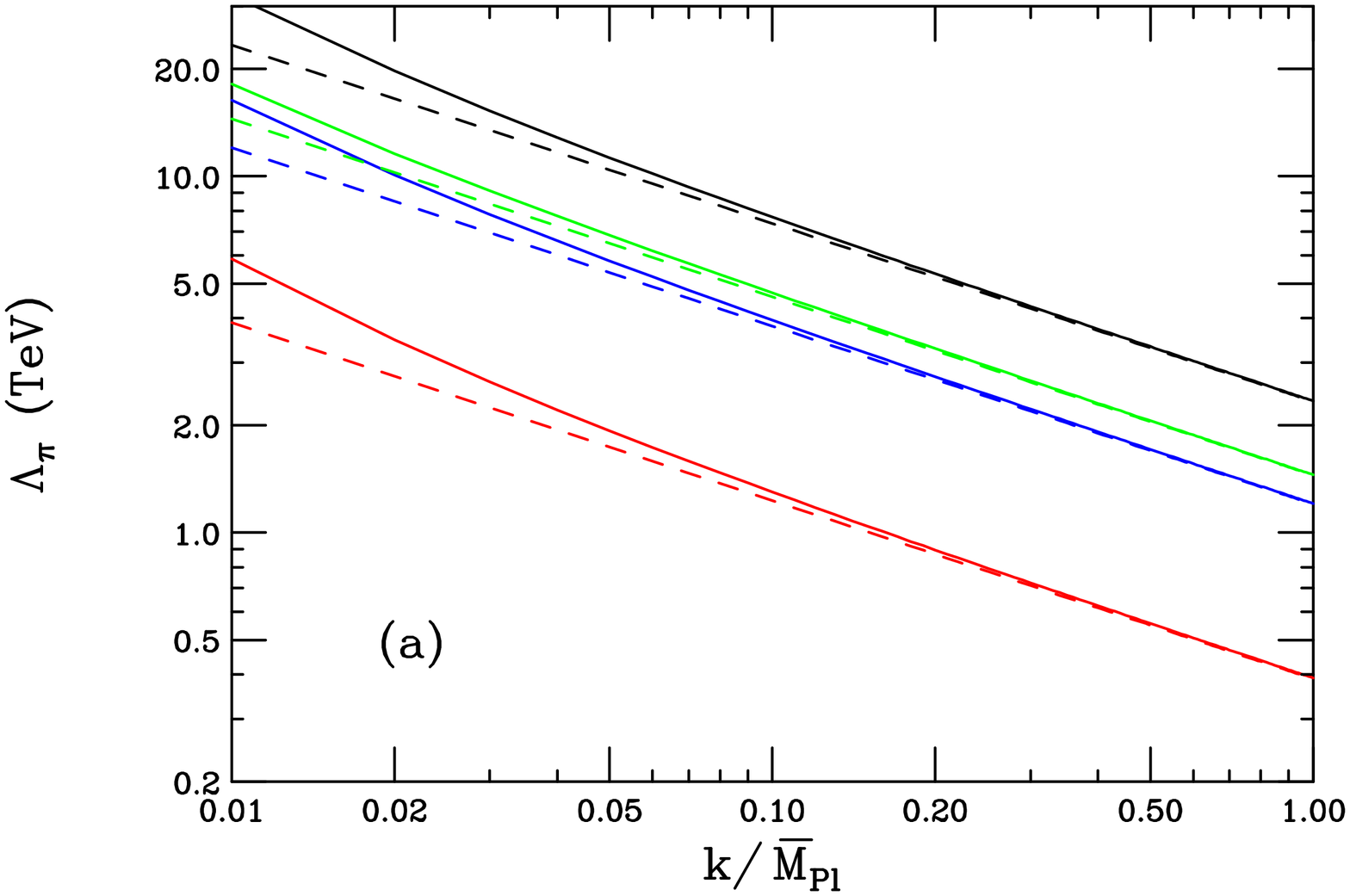,height=8.cm,width=12cm,angle=90}}
\vspace*{0.25cm}
\centerline{
\psfig{figure=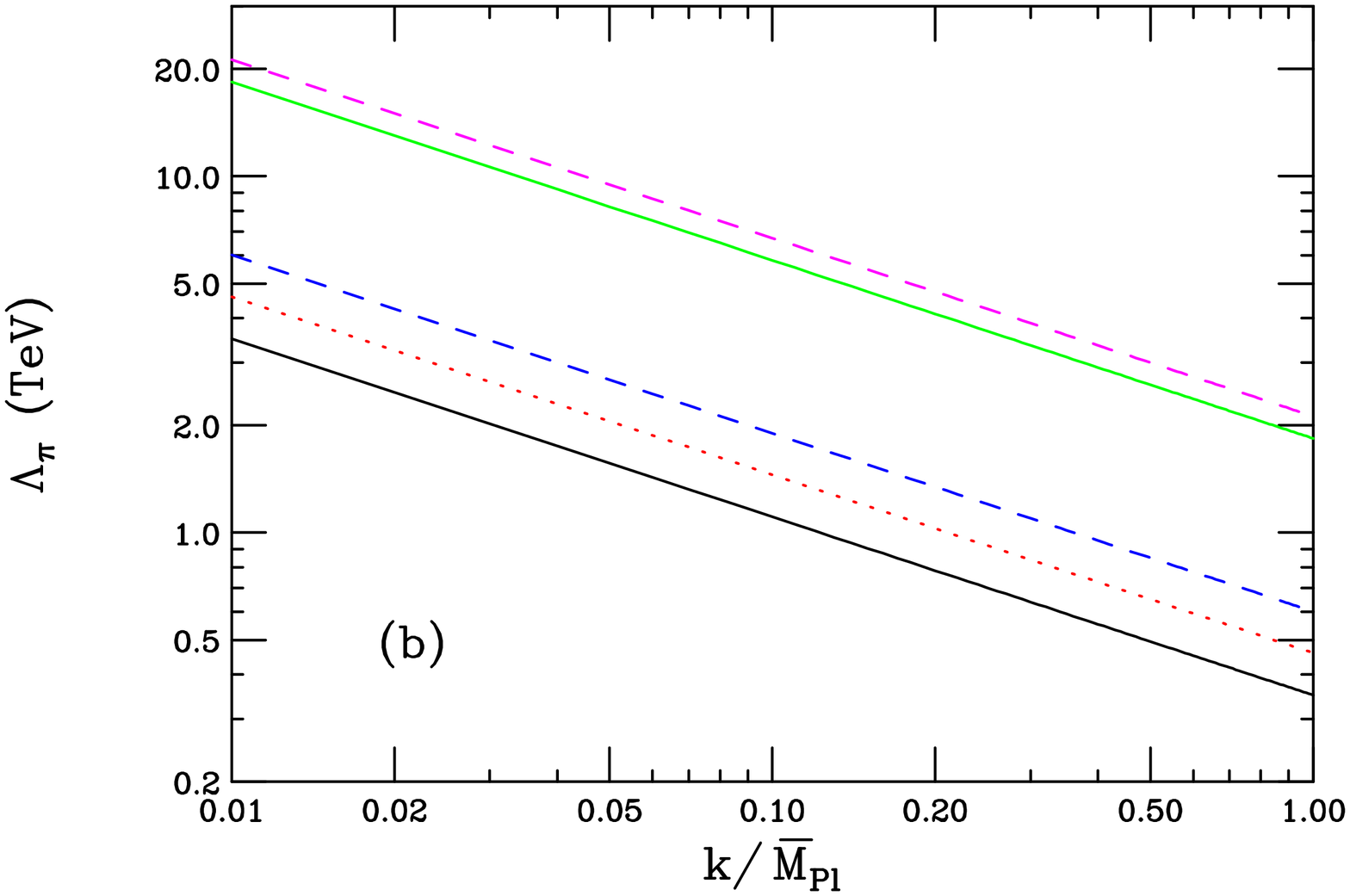,height=8.cm,width=12cm,angle=90}}
\vspace*{0.25cm}
\caption{Constraints in the $\Lambda_\pi-k/\mpl$ plane from virtual exchange
of the tower of KK gravitons.  The excluded region lies below the curves.
(a) The dashed curves assume the entire tower lies far above $\sqrt s$, while
the solid curves correspond to the case where the first two excitations are
close to $\sqrt s$.  From bottom to top the pairs of curves correspond to
LEP II at 195 GeV with 2.5 \infb\ of integrated luminosity; a linear 
\epem\ collider
at 500 GeV with 75 \infb; 500 GeV with 500 \infb; and 1 TeV with 200
\infb.  (b) From bottom to top the curves correspond to the Tevatron
Run I with 110 \inpb, Run II with 2 \infb, 
Run II with 30 \infb, the LHC with 10 \infb,
and 100 \infb.}
\label{cont}
\end{figure}

In the circumstance that gravitons are too massive to be directly produced at
colliders, their contributions to fermion pair production may still be felt
via virtual exchange.  For smaller values of $k/\mpl$, this would be similar 
to observing the effects of the SM $Z$ boson before the resonance turns on,
or for larger values, to searching for contact interactions.  The 4-fermion
matrix element is easily computed from the Lagrangian (\ref{effL}) and
is seen to reproduce that derived for the scenario of ADD 
with large extra factorizable dimensions\cite{me} with the replacement
\be
{\lambda\over M_s^4} \to {i^2\over 8\Lambda_\pi^2}\sum_{n=1}^\infty
{1\over s-m_n^2}\,.
\ee
The advantage in this scenario over the factorizable case is that there are
no divergences associated with performing the sum since there is
only one new dimension,
and hence uncertainties associated with the
introduction of a cut-off do not appear.  In the limit of $m_n^2\gg s$, the
sum over the KK graviton propagators becomes $[k\Lambda_\pi/\mpl]^{-2}
\sum_n 1/x_n^2$ which rapidly converges.  The $95\%$ C.L. search reach
in the $\Lambda_\pi - k/\mpl$ plane are given in Fig. \ref{cont} for 
various (a) \epem\
and (b) hadron colliders.  In \epem\ annihilation we have examined the 
unpolarized (and polarized for the case of high energy linear colliders) 
angular and $\tau$ polarization 
distributions, summing over $e,\mu,\tau,c,b$ (and $t$, if kinematically
accessible) final states, and
included initial state radiation, heavy quark tagging efficiencies, an
angular cut around the beam pipe, and $90\%$ beam polarization where
applicable. For hadron colliders we examined 
the lepton pair invariant mass
spectrum and forward-backward asymmetry in Drell-Yan production, for both
$e$ and $\mu$ final states.  We also investigated the case where the first
two excitations are too close to the collider center-of-mass energy
to use the approximation $m_n^2\gg s$.  The bounds in \epem\ annihilation
for this case are given by the solid curves in Fig. \ref{cont}(a).  We see
that there is very little difference in the resulting constraints.

\nn

As a last point, we note that whereas graviton tower emission was an
important probe of the ADD scenario, this is no longer true in the RS
model since the graviton states are so massive and can be individually
examined on resonance.

 In this paper we have explored the phenomenological implications of the 
Randall-Sundrum localized gravity model of non-factorizable 5-dimensional 
spacetime, and contrasted it with the ADD scenario.
We ($i$) derived the interaction of the KK tower of 
gravitons with the SM fields, ($ii$) obtained limits on the model parameters 
using existing data from colliders, both through direct production searches and 
via virtual exchange contributions, and estimated 
what future colliders can do to extend 
these bounds. ($iii$) We described the appearance of KK tower production 
at high energy linear colliders, the possible loss of the conventional 
KK signature of successive peaks due to 
the ever growing widths of these excitations, and ($iv$) demonstrated how 
measurements of the properties of the first KK state would completely determine 
the model parameters. 

We find the scenario of gravity localization 
to be theoretically very attractive, and even more
importantly, to have distinctive experimental tests.  We hope that future
experiment will eventually reveal the existence of higher dimensional
spacetime.

\noindent{\bf Acknowledgements}:

We would like to thank M. Dine, M. Schmaltz, and M. Wise for 
beneficial discussions.

%
\def\IJMP #1 #2 #3 {Int. J. Mod. Phys. A {\bf#1},\ #2 (#3)}
\def\MPL #1 #2 #3 {Mod. Phys. Lett. A {\bf#1},\ #2 (#3)}
\def\NPB #1 #2 #3 {Nucl. Phys. {\bf#1},\ #2 (#3)}
\def\PLBold #1 #2 #3 {Phys. Lett. {\bf#1},\ #2 (#3)}
\def\PLB #1 #2 #3 {Phys. Lett. B {\bf#1},\ #2 (#3)}
\def\PR #1 #2 #3 {Phys. Rep. {\bf#1},\ #2 (#3)}
\def\PRD #1 #2 #3 {Phys. Rev. D {\bf#1},\ #2 (#3)}
\def\PRL #1 #2 #3 {Phys. Rev. Lett. {\bf#1},\ #2 (#3)}
\def\PTT #1 #2 #3 {Prog. Theor. Phys. {\bf#1},\ #2 (#3)}
\def\RMP #1 #2 #3 {Rev. Mod. Phys. {\bf#1},\ #2 (#3)}
\def\ZPC #1 #2 #3 {Z. Phys. C {\bf#1},\ #2 (#3)}

\end{document}